\pdfoutput=1
\documentclass{article}
\usepackage{INTERSPEECH2022}
\usepackage{amsmath,graphicx,multirow,url,adjustbox}
\usepackage{spconf}
\usepackage{cite}
\usepackage{color,soul}
\usepackage{hyperref}
\usepackage{mathtools}
\usepackage{tabularx}
\usepackage{booktabs}
\usepackage{stfloats}
\usepackage[pdf]{graphviz}
\usepackage{pgfplots}
\pgfplotsset{width=\linewidth*1.04,compat=1.17}
\usepackage{subcaption}
\usepackage{listings}
\usepackage{url}
\definecolor{our_green}{HTML}{76b900}
\definecolor{our_blue}{HTML}{6d9eeb}

\usepackage{xpatch}
\makeatletter
\newcommand*{\addFileDependency}[1]{
  \typeout{(#1)}
  \@addtofilelist{#1}
  \IfFileExists{#1}{}{\typeout{No file #1.}}
}
\makeatother
\xpretocmd{\digraph}{\addFileDependency{#2.dot}}{}{}

\newcommand\blfootnote[1]{%
  \begingroup
  \renewcommand\thefootnote{}\footnote{#1}%
  \addtocounter{footnote}{-1}%
  \endgroup
}

\newcolumntype{Y}{>{\centering\arraybackslash}X}

\title{Multi-blank Transducers for Speech Recognition}

\name{\begin{tabular}{c} Hainan Xu$^1$, Fei Jia$^1$, Somshubra Majumdar$^1$, Shinji Watanabe$^2$, Boris Ginsburg$^1$
\end{tabular}
}
\address{
  $^1$NVIDIA, USA, \\
  $^2$Carnegie Mellon University, PA, USA \\
\small{\url{hainanx@nvidia.com}} }

\begin{document}
\maketitle

\begin{abstract}
This paper proposes a modification to RNN-Transducer (RNN-T) models for automatic speech recognition (ASR).
In standard RNN-T, the emission of a blank symbol consumes exactly one input frame; in our proposed method, we introduce additional blank symbols, which consume two or more input frames when emitted.
We refer to the added symbols as \emph{big blanks}, and the method \textit{multi-blank RNN-T}.
For training multi-blank RNN-Ts, we propose a novel \emph{logit under-normalization} method in order to prioritize emissions of big blanks.
With experiments on multiple languages and datasets, we show that multi-blank RNN-T methods could bring relative speedups of over +90\%/+139\% to model inference for English Librispeech and German Multilingual Librispeech datasets, respectively. The multi-blank RNN-T method also improves ASR accuracy consistently. We will release our implementation of the method in the NeMo (\url{https://github.com/NVIDIA/NeMo}) toolkit. 

\end{abstract}


\section{Introduction}
End-to-End (E2E) automatic speech recognition (ASR) systems can directly generate output sequences of text tokens from the input sequence of acoustic features, 
and have gradually surpassed hybrid ASR models \cite{povey2011kaldi} in terms of popularity and/or accuracy, and a lot of open-source toolkits \cite{watanabe2018espnet,ott2019fairseq,wang2019espresso,kuchaiev2019nemo,ravanelli2021speechbrain} are available. 
Among the common approaches, Attention-based Encoder and Decoder (AED) \cite{chorowski2015attention,chan2016listen}, Connectionist Temporal Classification (CTC)~\cite{graves2006connectionist} and Recurrent Neural Network Transducers (RNN-T)~\cite{graves2012sequence} are the most commonly used ones. 
Considerable research efforts have been spent in improving those ASR models, e.g. for computational efficiency \cite{li2019improving,kuang2022pruned,Ghodsi2020stateless,chen2016phone}, more flexible training scenarios \cite{pratap2022star,shinohara22_interspeech}, and different types of regularization methods \cite{xu2021regularizing,xu2021convolutional,yu2021fastemit}.
\blfootnote{Paper accepted at ICASSP 2023 conference.}

This paper focuses on RNN-T models\footnote{Although when first proposed, the term RNN-T was limited in meaning models that use LSTM encoders and decoders, we here refer to it as a general type of model that incorporates the transducer loss as its training objective.}.
An RNN-T model consists of an acoustic encoder (or simply, encoder), a decoder (also referred to as a prediction network or a label encoder), and a joint network. The acoustic encoder converts the input acoustic features into a higher-level representation; the decoder extracts the history context information at the label side; the joint network combines the output of the acoustic encoder and the decoder and outputs a probability distribution over the vocabulary. 
RNN-T achieves great performance  but it suffers from the slow inference speed and difficulty to train due to model structure and memory footprint\cite{mahadeokar2021alignment}. 
\cite{hu2020exploring} proposed using external alignment information to pre-train RNN-T to overcome the training difficulty and showed that pre-training can not only improve accuracy but also reduce the RNN-T model latency.
\cite{mahadeokar2021alignment} proposed using alignment to restrict RNN-Ts in streaming scenarios.
\cite{zhang2020transformer,han2020contextnet,gulati2020conformer,radfarconvrnn} all have proposed methods to improve the encoders for RNN-Ts.
The decoder, as well as the decoding algorithm of the RNN-T, have also been investigated, e.g. 
\cite{Ghodsi2020stateless,kim2020accelerating,kang2022fast}.

In this work, we focus on a relatively less investigated area of research on RNN-Ts -- the blank symbol and the loss function. 
Unlike standard RNN-Ts with a single blank symbol, we propose a multi-blank method, with additional blank symbols that explicitly model duration, and advance the input $t$ dimension by two or more frames. 
The method is straightforwardly implemented as an extension to standard RNN-Ts.
We show that the proposed multi-blank method improves WER and speeds up ASR inference consistently.
On Librispeech test sets, the method brings relative speedups of up to +92.9\% while also improving  WER. Besides English, we illustrate that the method also helps improve German ASR accuracy with up to +139.6\% speedup on the Multilingual LibriSpeech (MLS) dataset. 




\section{Multi-blank RNN-T}

\subsection{Blank symbol in RNN-T}
By leveraging a blank symbol in the model design, RNN-T does not need alignment information during training.
In the RNN-T framework, a label sequence could be augmented by adding an arbitrary number of blanks at any position of the sequence, and during RNN-T model training, for any input sequence, it tries to maximize the probability sum over all augmented sequences of the correct labels. In Figure~\ref{fig:trellis_ori}, we demonstrate an output probability lattice of standard RNN-T model by following~\cite{graves2012sequence}. 
The probability of  observing the first $u$ output sequence elements in the first $t$ transcription sequence is represented by node $(t, u)$.
An upward pointing arrow leaving node $(t,u)$ represents $y(t,u)$, the probability of outputting an actual label; and a rightward pointing arrow represents $\O(t,u)$, the probability of outputting a blank at $(t,u)$. 
Note that when outputting an actual label, $u$ would be incremented by one; and when a blank is emitted, $t$ is incremented by one.

During inference, an RNN-T model emits at least one token per input frame, and produces a sequence of non-blank and blank symbols. Typical RNN-T output looks like this,
\fbox{$\O$ $\O$ $\O$  $\O$  $\O$ $\O$ \_how  $\O$  $\O$  $\O$ $\O$ $\O$ \_are  $\O$  $\O$  $\O$ $\O$ \_you  $\O$  $\O$ $\O$ $\O$}
where  $\O$ is the blank symbol. Those blank symbols are omitted in post-processing in order to generate the final ASR outputs. 

\subsection{Multi-blank RNN-T}
With the example given in the last section, we note that empirically, a typical RNN-T model generates  more blank symbols than non-blanks during inference, which means the model spends a lot of computation generating labels that are not going to be in the final outputs.
In this work, we propose \emph{multi-blank} RNN-Ts, which not only use the standard blanks like standard RNN-T, but also
introduces \emph{big blank} symbols. Those big blank symbols could be thought of as blank symbols with explicitly defined \emph{duration}s -- once emitted, the big blank advances the $t$ by more than one, e.g. two or three.
A multi-blank model could use an arbitrary number of blanks with different durations, represented as a set $\mathcal{N}$ containing all
possible blank durations. We require $1 \in \mathcal{N}$.
Note, standard RNN-Ts could be seen as a special case of multi-blank RNN-Ts where $\mathcal{N} = \{1\}$.

\begin{figure}[tb]
\begin{subfigure}[b]{0.98\linewidth}
   \centering
  \centerline{\includegraphics[scale=0.25]{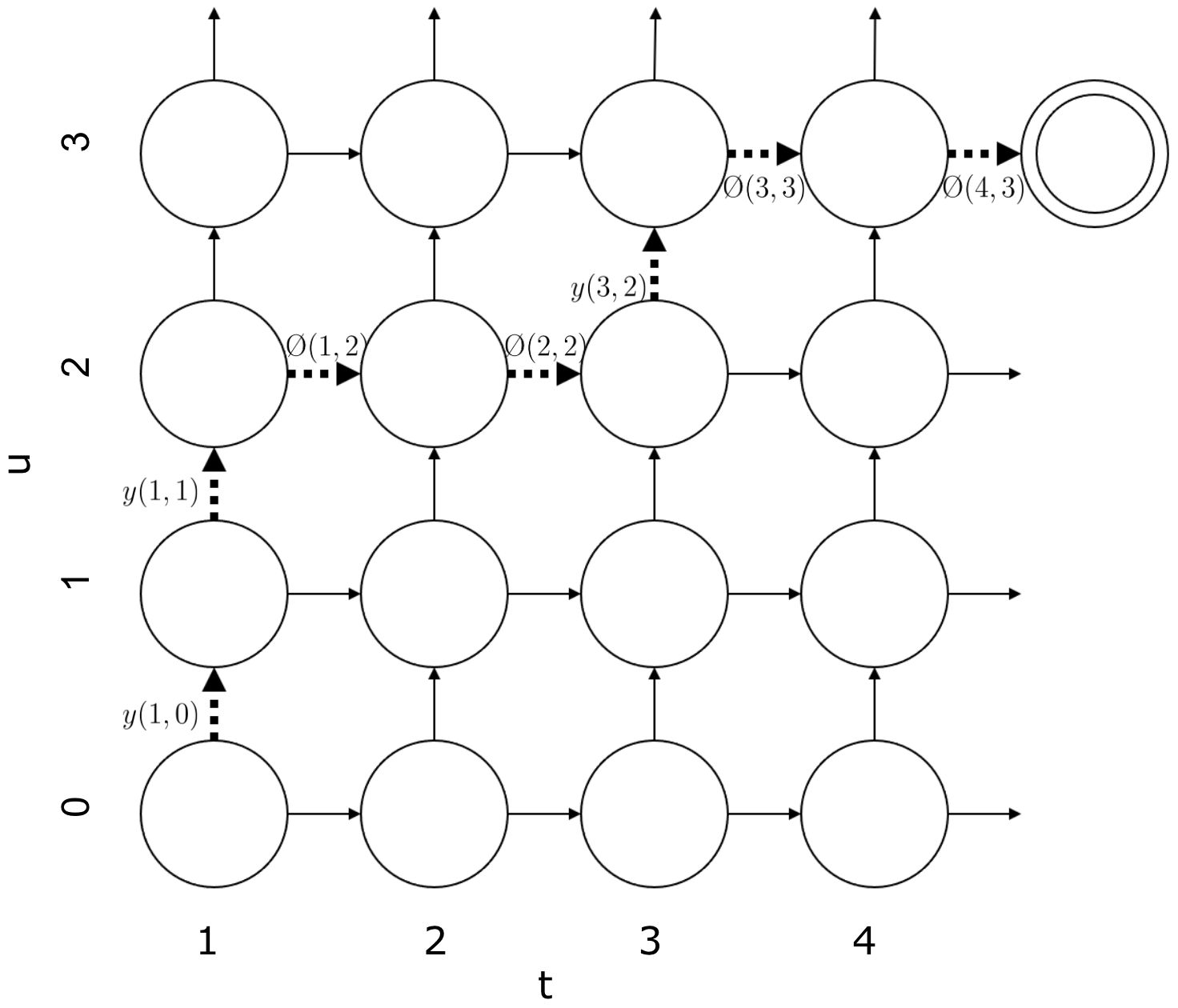}}
    \caption{\emph{Output probability lattice of standard RNN-T. The probability of observing the first $u$ output labels  in the first $t$  frames is represented by node $(t, u)$. An upward pointing arrow leaving node $(t,u)$ represents $y(t,u)$, the probability of outputting an actual label; and a rightward pointing arrow represents $\O(t,u)$, the probability of outputting a blank at $(t,u)$. } }
  \label{fig:trellis_ori}
\end{subfigure}
\begin{subfigure}[b]{0.98\linewidth}
   \centering
  \centerline{\includegraphics[scale=0.25]{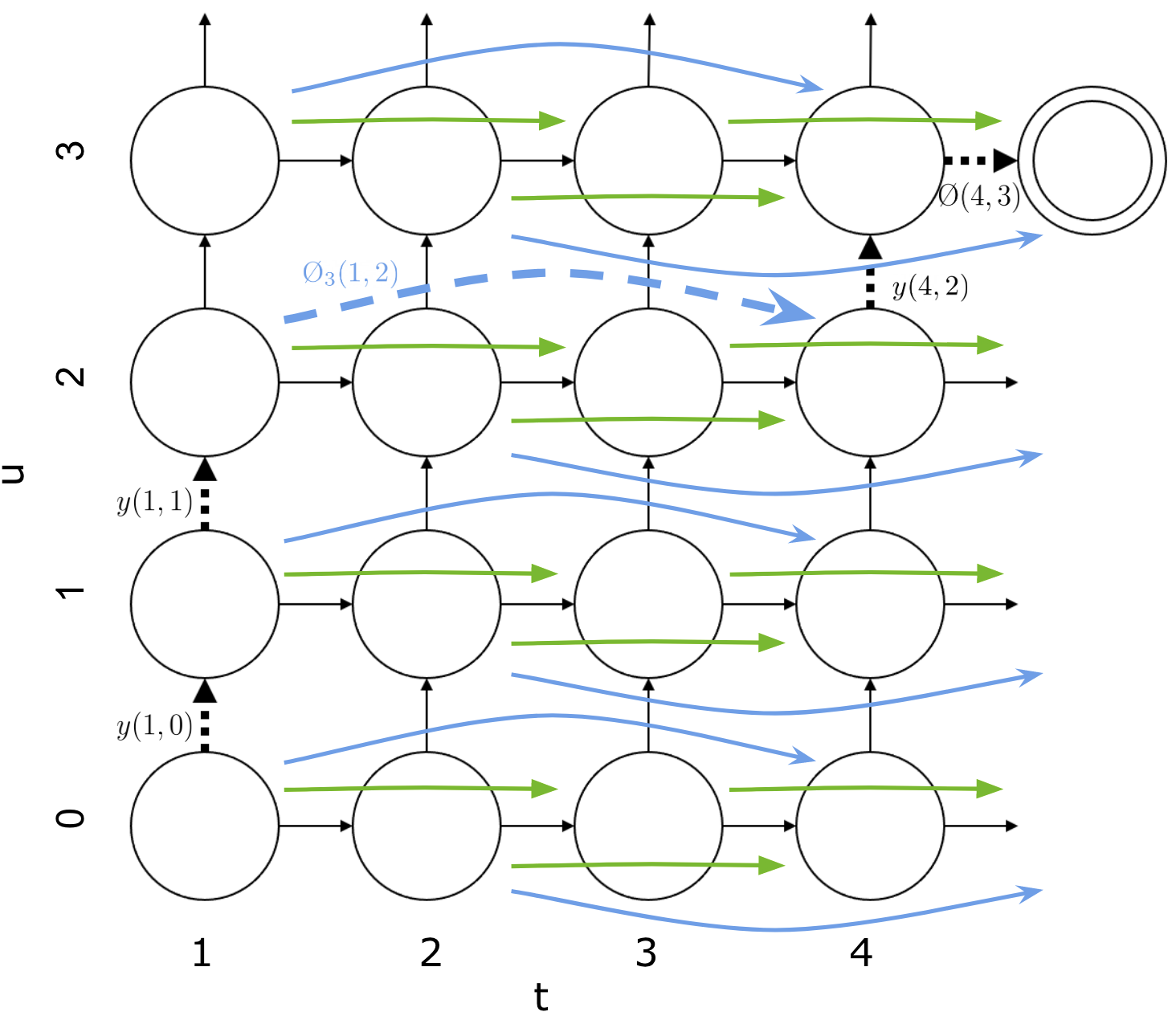}}
  \caption{\emph{Output probability lattice with multi-blank RNNT. {\color{our_green} Green} arcs here denote big blanks with duration $2$, represented as $\O_2(t,u)$ and {\color{our_blue} blue}  denotes big blanks with duration $3$, represented as $\O_3(t,u)$. The dashed line is a possible path with the proposed method.}}
  \label{fig:trellis_our}
\end{subfigure}
\caption{Output probability lattice of a standard RNN-T model as in~\cite{graves2012sequence} and of a multi-blank RNNT.
}
\label{fig:trellis}
\end{figure}

We compare the probability lattices of standard RNN-T and multi-blank RNN-T in Figures \ref{fig:trellis}.
Figure~\ref{fig:trellis_ori} is similar to the original Figure from \cite{graves2012sequence}, which has only one standard blank symbol with duration 1 ($\mathcal{N}=\{1\}$);
Figure~\ref{fig:trellis_our} uses  two big blanks with durations $m=2$ and $m=3$, thus $\mathcal{N}=\{1, 2, 3\}$. The transition arcs corresponding to big blanks are in colors {\color{our_green}green} and {\color{our_blue}blue}.




\subsection{Forward-backward algorithm}
To enable multi-blank RNN-T, the forward-backward algorithm needs to be modified. We follow the notations of \cite{graves2012sequence} for describing the forward-backward algorithm. 
With the standard RNN-T models, the forward weights ($\alpha$)
and backward weights ($\beta$) are computed as
\begin{equation}
\begin{split}
    \alpha(t,u)& = \alpha(t,u - 1)y(t,u - 1) + \alpha(t - 1,u)\O(t-1,u),\\
    \beta(t, u)& = \beta(t, u + 1) y(t, u)+ \beta(t + 1, u)\O(t, u) .
\end{split}
\end{equation}
For multi-blank RNN-Ts, with a predefined $\mathcal{N}$,
we have
\begin{equation}
\begin{split}
 \alpha(t, u)=&  \alpha(t, u - 1) y(t, u - 1) + \sum_{m \in \mathcal{N}}\alpha(t-m, u) \O_{m}(t-m, u), \\
    \beta(t, u) = &\beta(t, u + 1) y(t, u) + \sum_{m \in \mathcal{N}} \beta(t + m, u) \O_m(t, u),\\
\end{split}
\end{equation}
where $\O_m(t,u)$ represent the probability of a big blank with duration $m$ at lattice location $(t,u)$.
\footnote{
We remind the readers that  like standard RNN-Ts, we need special attention to the boundary cases when $t<m$ for the forward weights or $t + m \geq T$ for the backward weights, where big blank transitions don't exist. In those cases, the big blank weight should be omitted.}


\subsection{Model inference}\label{inference}
In standard decoding algorithms for RNN-Ts, the emission of a blank symbol advances input by one frame.
Naturally, for inference with multi-blank RNN-T models, we need to change the behavior of the decoding algorithm for big blank emissions.
With the multi-blank models, when a big blank with duration $m$ is emitted, the decoding loop increments $t$  by exactly $m$.
This allows the inference to skip  frames and thus become faster.

\section{Logits Under-normalization}
Since emissions of big blanks could bring speedup to inference, we want to prioritize emissions of big blanks.
We propose a modified RNN-T loss function for this purpose.
In standard RNN-Ts,
the $y(t, u)$ and $\O(t, u)$ terms are probabilities of the correct label or blank emitted at  $(t, u)$ location in the probability lattice.
This is typically implemented with a log\_softmax function call, so that the logits represent log probabilities
\footnote{Note this is just to better illustrate the method. In practice, we adopt the \emph{function merging} method from \cite{li2019improving} which  performs the normalization and summation together. We will release the code upon acceptance of this paper.}:
\begin{equation}
    \text{logits} =
    \text{log\_softmax}(y).
\end{equation}
Here we  \emph{under-normalize} the logits by adding an extra term.
\begin{equation}\label{logits}
\text{logits} = 
\text{log\_softmax}(y) - \sigma,
\end{equation}
where $\sigma$ is chosen to be 0.05 in our experiments. 
In the modified RNN-T computation, the weight of a complete path $\pi$ would be the sum of those under-normalized logits, computed as,
\begin{equation}
\text{weight}(\pi) = \log P(\pi) - \sigma \cdot |\pi|,
\end{equation}
where log P($\pi$) comes from log\_softmax($y$), and $|\pi|$ represents the number of emissions (total number of labels and any types of blanks) in the path $\pi$.
Note that RNN-T loss requires summing over the weights of all paths\cite{graves2012sequence}. 
 With the added terms, the loss does not sum over the probabilities \emph{uniformly}, but applies a weight depending on $|\pi|$, i.e.
\begin{equation}
\begin{split}
    \mathcal{L}_\text{multi-blank RNN-T} &= \log \sum_\pi  \exp (\text{weight($\pi$)}) \\
    & = \log \sum_\pi \frac{P(\pi)} {\exp(\sigma \cdot |\pi| )}.
\end{split}
\end{equation}
The added weight penalizes longer paths, and therefore  would prioritize the emission of blanks with larger durations since they cover multiple frames and make the path shorter. 
Note that under-normalization has no effect on the original RNN-T  since all paths are of the same length and thus penalized equally.



\section{Experiments}\label{experiments}
We evaluate our methods with a Conformer-RNN-T model with a stateless decoder. We find that the stateless-decoder models consistently outperform  LSTM-decoder models both in terms of accuracy and speed. We actually have done extensive experiments with RNN-Ts with standard LSTM-decoders as well and all our conclusions about the multi-blank methods still hold. 

Our model extracts acoustic features with frame-rate 10ms and window-size 25ms. The Conformer encoder has a convolution layer at the beginning of the network that performs subsampling on the input.
The stateless decoders use the concatenation of  embeddings of the last two context words as the output. 
The models use \emph{byte-pair encoding} \cite{sennrich2015neural} as the text representation, and vocabulary sizes are chosen to be 1024. 

Although theoretically multi-blank RNN-T models require slightly increased computation during training, we observe negligible increases in training time compared to standard RNN-Ts.
Therefore, for all datasets, we report WER and the decoding time in seconds 
with non-batched greedy inference. We also include the relative speedup factor, represented in percentage, for different types of models compared to the corresponding baseline, which uses standard blank only with $\mathcal{N}=\{1\}$.
For all multi-blank experiments, we use $\sigma = 0.05$.
We did minimal tuning with this parameter as our initial experiments indicate the results aren't sensitive to this value.



\subsection{Librispeech results}
Our Librispeech models are trained with the full Librispeech dataset\cite{panayotov2015librispeech}, augmented 3-times using speed perturbation factors of 0.9x 1.0x, 1.1x. We use the \emph{conformer-rnnt-large} configuration in NeMo\footnote{See \url{examples/asr/conf/conformer/conformer_transducer_bpe.yaml} in NeMo repository.}, which has around 120M parameters. 
To figure out the optimal subsampling rates, we first conduct experiments shown in Table \ref{lib_4x_8x}. We see that 4X subsampling gives the best accuracy,
which we choose as our baseline for comparisons. Since our feature extraction uses a frame-rate of 10ms, this means the equivalent frame-rate used in each decoding step is 40ms.


\begin{table}
    \centering
    \begin{tabular}{ccc}
        \toprule
        \multirow{2}{*}{ subsampling} & \multicolumn{2}{c}{LS test-other} \\
    \cmidrule(lr){2-3}     
        &  WER ($\%$) & time (sec)\\
     \midrule
       4x   & 5.43  & 243\\
     \midrule
      8x   & 5.56 & 150\\
      \midrule
      16x & 6.38 & 106\\
     \bottomrule
    \end{tabular}
    \caption{Baseline Conformer-RNNT (Large) with different sub-sampling rates in the encoder: WER ($\%$), and inference time (seconds) for Librispeech test-other. The inference time is computed on the whole test-other dataset with batch=1}
    \label{lib_4x_8x}
\end{table}



The comparisons with multi-blank methods are shown in Table \ref{fig:librispeech_stateless}. 
From the results, 
we see that in all cases, multi-blank RNN-T models achieve better WERs than standard RNN-T, and their inference is faster. 
More speedup is seen with models using big blanks with larger durations. We see speedups over 90\% in two of our models that use big blanks with duration up to 8. 
Here we  point out that our best models are able to achieve  inference speed between that of  8X  and 16X models in Figure~\ref{lib_4x_8x}. Moreover, unlike higher sub-sampling rates that hurt WERs, our methods give better WER numbers.
Readers are reminded that for models with the 8X and 16X subsampling, the subsampling at the beginning of the encoder directly impacts the computational complexity of the later self-attention operations, hence part of the speedup actually comes from reduced encoder computation, while in our approach all speedup comes from the decoding loops.

\begin{table}
    \centering
    \begin{tabular}{ ccccc  }
    \toprule
     {multi-blank } &  \multicolumn{3}{c}{LS test-other}  \\
    \cmidrule(lr){2-4} 
     config $\mathcal{N}$  & WER ($\%$) & time (sec)  & speedup ($\%$) \\
    \midrule

     \multicolumn{1}{c}{
     baseline}  &5.43 & 243 & - \\
    \{1, 2\} &  5.24 & 174 & 39.7 \\
    \{1, 2, 4\} &  5.32 & 140 & 73.6 & \\
    \{1, 2, 4, 8\} &  5.37 & 126 & 92.9 \\
    \{1,2,3,4,5,6,7,8\}&    5.27 & 126 & 92.9 \\

    \bottomrule
    \end{tabular}
    \caption{Librispeech test-other: WER(\%), inference time (seconds), and relative speedup for RNN-T and different multi-blank configurations. Inference time is computed with batch=1. Multi-blank config $\mathcal{N}$ is the  set of blank durations the model supports, including the standard blank. For example, \{1,2,4\} means the model has a standard blank (with duration 1), and big blanks with durations 2 and 4.}
    \label{fig:librispeech_stateless}
\end{table}

\subsection{German ASR results}

For German experiments, we use our internal German dataset which consists of around 2070 hours of audio data and 813,000 utterances. 
We use RNN-T models with stateless decoders identical to the Librispeech models reported in the previous section and report results on the German test sets in VoxPopuli \cite{wang2021voxpopuli} and Multilingual Librispeech (MLS) \cite{pratap2020mls}.
The results are shown in Table \ref{fig:german_stateless_4x}.
We see similar trends compared to Librispeech models. Multi-blank methods bring speedups up to around 90\% for the VoxPopuli dataset, and around 140\% on MLS, while improving ASR accuracy when $\mathcal{N} = \{1,2,4,8\}$. Consistent accuracy improvements and various speedup factors are seen with other configurations as well, with a maximum WER gain of over 1\% absolute, when $\mathcal{N} = \{1,2,4\}.$




\begin{table}
    \centering
    \begin{tabular}{ccccccc}\toprule
    {multi-blank }&  \multicolumn{3}{c}{German VoxPopuli} & \multicolumn{3}{c}{German MLS} \\
    \cmidrule(lr){2-4} \cmidrule(lr){5-7}
     config $\mathcal{N}$  & WER & time  & speedup ($\%$) & WER & time &  speedup ($\%$)\\
    \midrule

     \multicolumn{1}{c}{baseline} & 8.67 & 230 & - & 4.00 & 544 & - \\

    \{1,2\}   & 8.27 & 166 & 38.6 & 3.86 & 363 & 49.9 \\
    \{1,2,4\} & 7.66 & 134 & 71.6 & 3.96 & 272 & 100.0 \\
    \{1,2,4,8\} & 7.89 & 122 & 88.5 & 3.89 & 227 & 139.6 \\
    \{1,2,...,7,8\}& 7.82 & 121 & 90.1 &  3.85 & 230 & 136.5  \\
    \bottomrule
    \end{tabular}
    \caption{German ASR results: WER(\%), inference time (seconds), and relative speedup for standard and different multi-blank RNN-Ts. Inference time is computed with batch=1. See Table \ref{fig:librispeech_stateless} for explanation  of multi-blank config $\mathcal{N}$.}
    \label{fig:german_stateless_4x}
\end{table}




\section{Analysis}

\subsection{Impact of under-normalization for training}
We  compare our Librispeech models' ASR accuracy and inference speed on test-other in Table \ref{under}, with our without logit under-normalization. We see that without under-normalization, smaller speedup could be seen, but models with big blanks of long durations don't necessarily run faster, e.g. \{1,2,4,8\} is actually slower than \{1,2,4\}. However, with under-normalization, much bigger speedup factors could be achieved without significant changes in ASR accuracy, and we consistently see bigger speedup when using big blanks with longer durations.

\begin{table}[t]
    \centering
    \begin{tabular}{ cccccccc  }
    \toprule
     {multi-blank} &  \multicolumn{3}{c}{normal soft-max}  & \multicolumn{3}{c}{logits under-norm}  \\
    \cmidrule(lr){2-4} \cmidrule(lr){5-7}
     config $\mathcal{N}$  & WER & time  & speedup ($\%$)& WER & time &  speedup ($\%$)\\
    \midrule
    baseline & 5.43 & 243 & - \\
    \{1,2\} & 5.37 & 212 & 14.6 & 5.24 & 174 & 39.7 \\
    \{1,2,4\} & 5.21 & 195 & 24.6 & 5.32 & 140 & 73.6  \\
    \{1,2,4,8\} & 5.28 & 196 & 24.0 & 5.37 & 126 & 92.9 \\
    \{1,2,...,7,8\} & 5.18 & 192 & 26.6 & 5.27 & 126 & 92.9 & \\

    \bottomrule
    \end{tabular}
    \caption{WER(\%), inference time(sec) and relative speedup(\%), on Librispeech test-other, with/without logits under-normalization during training. See Table \ref{fig:librispeech_stateless} for explanation  of multi-blank config $\mathcal{N}$.}
    \label{under}
\end{table}

\subsection{Big blank emission frequency}
\begin{figure}[h]
    \centering
    \includegraphics[scale=0.3]{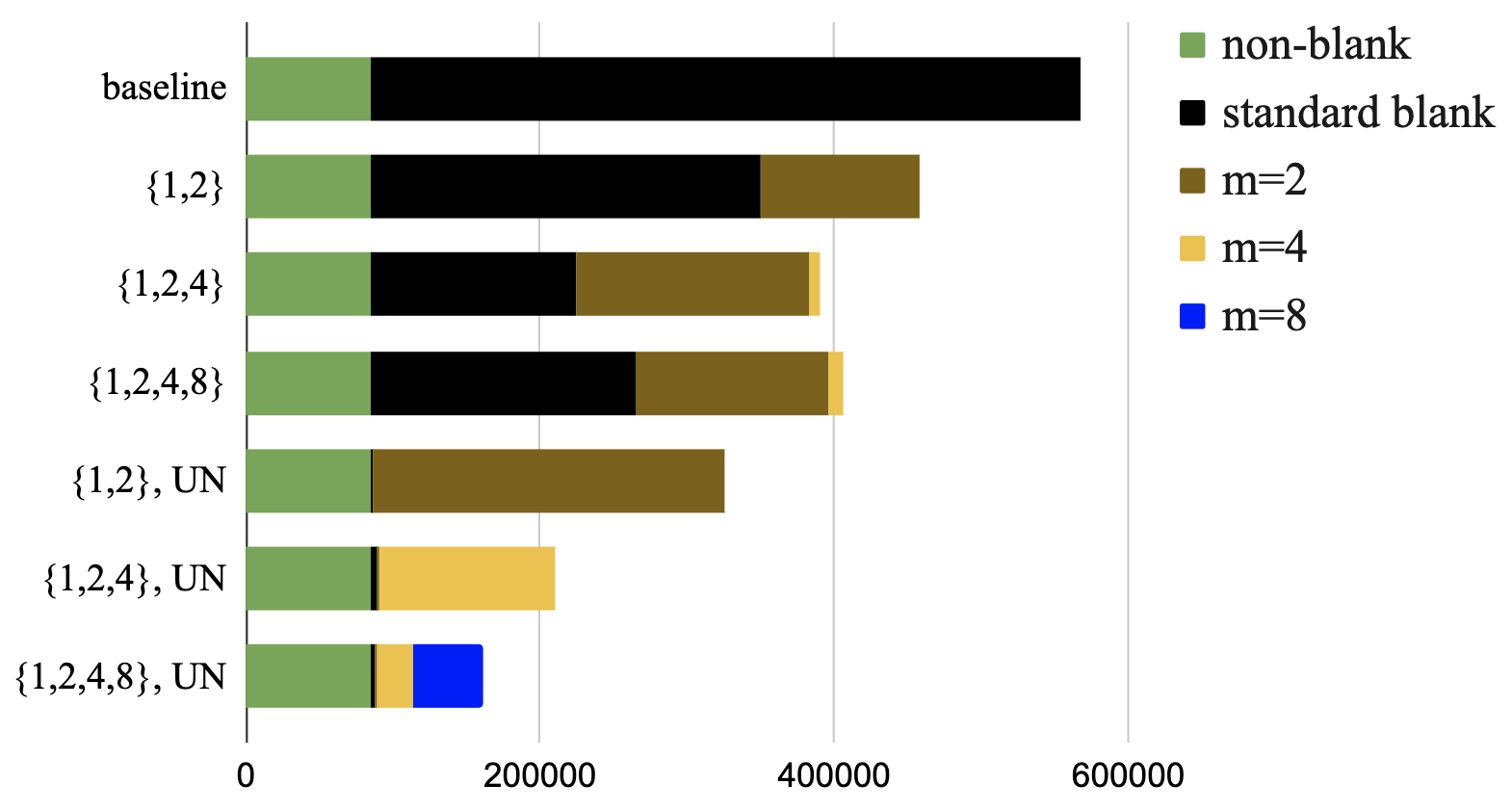}
    \caption{Emission distribution on Librispeech test-other. UN means the model is trained with logits under-normalization.}
    \label{blank_frequency}
\end{figure}

We study the distribution of model emissions during  inference, shown in
Figure~\ref{blank_frequency}, where we plot the number of emission counts for different types of symbols with different models on Librispeech test-other. The model is marked with ``UN'' when it is trained with logits under-normalization. 
We can see without under-normalization, standard blank emission is suppressed a bit, but longer big blanks are not frequently emitted; however, when under-normalization is used during training, standard blank emissions are drastically decreased, and longer blanks appear significantly more frequently. Readers are reminded that the total length of the bars for each model represents the total number of emissions, which also equals the number of decoding steps during inference. This explains the more significant speedup for models trained with under-normalization.

\subsection{Efficient batched inference for multi-blank transducers}
The multi-blank RNN-T method is ideal for on-device speech recognition in that it brings significant speedup in recognition speed as well as better accuracy. With that being said, the method also supports batched inference to run on the server side.
In Section \ref{inference}, we mentioned that during inference, when a big blank symbol is emitted, it should advance the input $t$  by the duration corresponding to the big blank. This makes it non-trivial to implement exact batched inference for multi-blank RNN-Ts, since different utterances in the same batch might output blanks with different durations, making it hard to fully parallelize the computation.

We propose an \emph{inexact} batched inference method: if different utterances in a batch emit blanks of different durations, we increment $t$  by the minimum of those durations, e.g. if at frame $t$, different utterances emit blanks of durations 1, 2, and 4, then the $t$ is incremented by 1 for the next step in the batched decoding. This  allows for better parallelization for different utterances in the same batch.

We run the proposed method for batched inference with multi-blank RNN-T models ($	\mathcal{N}=\{1,2,4,8\}$) on the Librispeech test-other dataset and report the results in Table \ref{batch}. 
We see with the baseline, larger batch-sizes speed up model inference with diminishing returns, since large batch-sizes would result in more wasted computation due to padding; for multi-blank models, when the batch-size goes from 1 to 8, the improved parallelism brings speedup in inference; however, after that, it becomes slower again. Besides the padding factors, this is also because with larger batches, it is less likely for all utterances to emit big blanks of large durations, and thus it has to perform more decoding steps by picking the minimum of those durations.
Also, we notice slight differences in WER with different batch-sizes. This is because this batched inference method is not equivalent to running utterances in a non-batched mode (hence the method is inexact), in that
when a big blank is emitted for an utterance, the decoding process  might not advance the $t$ exactly according to the predicted duration, but pick a smaller duration instead, and this might result in  small perturbations in the ASR outputs.

\begin{table}[]
    \centering
    \begin{tabular}{cccccc}
    \toprule
 \multirow{2}{*}{batch-size} & \multicolumn{2}{c}{baseline}  & \multicolumn{3}{c}{multi-blank} \\
  \cmidrule(lr){2-3} \cmidrule(lr){4-6}
         & WER(\%) & time (sec) & WER(\%) & time (sec) & speedup (\%)\\
        \midrule
        1  & 5.43 & 243 & 5.37 & 126 & 92.9  \\
        2  & 5.43 & 169 & 5.40 & 90 & 87.8 \\
        4  & 5.43 & 134 & 5.40 & 78 & 71.8 \\
        8  & 5.43 & 114 & 5.37 & 77 & 48.1 \\
        16 & 5.43 & 103 & 5.35 & 80 & 28.8 \\
         \bottomrule
    \end{tabular}
    \caption{WER(\%) and inference time (sec) on Librispeech test-other with different batch-sizes. For multi-blank models, when different utterances in a batch emit blanks of different durations, $t$ is advanced by the minimum of those durations, making this an \emph{inexact} inference method, resulting in small differences in WERs. Speedup is computed relative to the baseline model with the same batch-size.}
    \label{batch}
\end{table}

\section{Conclusion and Future Work}
In this paper, we propose \emph{multi-blank RNN-T models}, which include standard blanks and also \emph{big blanks}, whose emission consumes multiple frames.
We show with multiple datasets in different languages that this design helps speed up the model inference and improves ASR accuracy.
In order to prioritize the emission of big blanks, we proposed a special RNN-T training method, which performs under-normalization of the logits before RNN-T computation, and brings further speedup to inference.
Our best models  bring between +90\% to +140\% relative speedup for inference on different datasets, while achieving better ASR accuracy. 
For future work, we will work on implementing beam-search for multi-blank RNNTs as well as evaluating the models in streaming modes; we will also perform analysis to investigate exactly why multi-blank models give better WERs than standard RNNT models.

\bibliography{refs}

\begin{thebibliography}{10}
\providecommand{\url}[1]{#1}
\csname url@samestyle\endcsname
\providecommand{\newblock}{\relax}
\providecommand{\bibinfo}[2]{#2}
\providecommand{\BIBentrySTDinterwordspacing}{\spaceskip=0pt\relax}
\providecommand{\BIBentryALTinterwordstretchfactor}{4}
\providecommand{\BIBentryALTinterwordspacing}{\spaceskip=\fontdimen2\font plus
\BIBentryALTinterwordstretchfactor\fontdimen3\font minus
  \fontdimen4\font\relax}
\providecommand{\BIBforeignlanguage}[2]{{%
\expandafter\ifx\csname l@#1\endcsname\relax
\typeout{** WARNING: IEEEtran.bst: No hyphenation pattern has been}%
\typeout{** loaded for the language `#1'. Using the pattern for}%
\typeout{** the default language instead.}%
\else
\language=\csname l@#1\endcsname
\fi
#2}}
\providecommand{\BIBdecl}{\relax}
\BIBdecl

\bibitem{povey2011kaldi}
D.~Povey, A.~Ghoshal, G.~Boulianne, L.~Burget, O.~Glembek, N.~Goel,
  M.~Hannemann, P.~Motlicek, Y.~Qian, P.~Schwarz, J.~Silovsky, G.~Stemmer, and
  K.~Vesely, ``The {Kaldi} speech recognition toolkit,'' in \emph{Workshop on
  Automatic Speech Recognition and Understanding}, 2011.

\bibitem{watanabe2018espnet}
S.~Watanabe, T.~Hori, S.~Karita, T.~Hayashi, J.~Nishitoba, Y.~Unno, N.~{Enrique
  Yalta Soplin}, J.~Heymann, M.~Wiesner, N.~Chen, A.~Renduchintala, and
  T.~Ochiai, ``{ESPnet}: End-to-end speech processing toolkit,'' in
  \emph{Interspeech}, 2018.

\bibitem{ott2019fairseq}
M.~Ott, S.~Edunov, A.~Baevski, A.~Fan, S.~Gross, N.~Ng, D.~Grangier, and
  M.~Auli, ``Fairseq: A fast, extensible toolkit for sequence modeling,'' in
  \emph{Proceedings of NAACL-HLT 2019: Demonstrations}, 2019.

\bibitem{wang2019espresso}
Y.~Wang, T.~Chen, H.~Xu, S.~Ding, H.~Lv, Y.~Shao, N.~Peng, L.~Xie, S.~Watanabe,
  and S.~Khudanpur, ``Espresso: A fast end-to-end neural speech recognition
  toolkit,'' in \emph{Automatic Speech Recognition and Understanding Workshop
  (ASRU)}, 2019.

\bibitem{kuchaiev2019nemo}
O.~Kuchaiev, J.~Li, H.~Nguyen, O.~Hrinchuk, R.~Leary, B.~Ginsburg, S.~Kriman,
  S.~Beliaev, V.~Lavrukhin, J.~Cook \emph{et~al.}, ``Nemo: a toolkit for
  building {AI} applications using neural modules,'' \emph{arXiv:1909.09577},
  2019.

\bibitem{ravanelli2021speechbrain}
M.~Ravanelli, T.~Parcollet, P.~Plantinga, A.~Rouhe, S.~Cornell, L.~Lugosch,
  C.~Subakan, N.~Dawalatabad, A.~Heba, J.~Zhong \emph{et~al.}, ``{SpeechBrain}:
  A general-purpose speech toolkit,'' \emph{arXiv:2106.04624}, 2021.

\bibitem{chorowski2015attention}
J.~K. Chorowski, D.~Bahdanau, D.~Serdyuk, K.~Cho, and Y.~Bengio,
  ``Attention-based models for speech recognition,'' \emph{Advances in neural
  information processing systems}, vol.~28, 2015.

\bibitem{chan2016listen}
W.~Chan, N.~Jaitly, Q.~Le, and O.~Vinyals, ``Listen, attend and spell: A neural
  network for large vocabulary conversational speech recognition,'' in
  \emph{ICASSP}, 2016.

\bibitem{graves2006connectionist}
A.~Graves, S.~Fern{\'a}ndez, F.~Gomez, and J.~Schmidhuber, ``Connectionist
  temporal classification: labelling unsegmented sequence data with recurrent
  neural networks,'' in \emph{ICML}, 2006.

\bibitem{graves2012sequence}
A.~Graves, ``Sequence transduction with recurrent neural networks,''
  \emph{arXiv:1211.3711}, 2012.

\bibitem{li2019improving}
J.~Li, R.~Zhao, H.~Hu, and Y.~Gong, ``Improving {RNN} transducer modeling for
  end-to-end speech recognition,'' in \emph{Automatic Speech Recognition and
  Understanding Workshop (ASRU)}, 2019.

\bibitem{kuang2022pruned}
F.~Kuang, L.~Guo, W.~Kang, L.~Lin, M.~Luo, Z.~Yao, and D.~Povey, ``Pruned
  {RNN-T} for fast, memory-efficient {ASR} training,'' \emph{arXiv:2206.13236},
  2022.

\bibitem{Ghodsi2020stateless}
M.~Ghodsi, X.~Liu, J.~Apfel, R.~Cabrera, and E.~Weinstein, ``{RNN-Transducer}
  with stateless prediction network,'' in \emph{ICASSP}, 2020.

\bibitem{chen2016phone}
Z.~Chen, W.~Deng, T.~Xu, and K.~Yu, ``Phone synchronous decoding with {CTC}
  lattice.'' in \emph{Interspeech}, 2016, pp. 1923--1927.

\bibitem{pratap2022star}
V.~Pratap, A.~Hannun, G.~Synnaeve, and R.~Collobert, ``Star temporal
  classification: Sequence classification with partially labeled data,''
  \emph{arXiv:2201.12208}, 2022.

\bibitem{shinohara22_interspeech}
Y.~Shinohara and S.~Watanabe, ``{Minimum latency training of sequence
  transducers for streaming end-to-end speech recognition},'' in \emph{Proc.
  Interspeech 2022}, 2022, pp. 2098--2102.

\bibitem{xu2021regularizing}
H.~Xu, K.~Audhkhasi, Y.~Huang, J.~Emond, and B.~Ramabhadran, ``Regularizing
  word segmentation by creating misspellings.'' in \emph{Interspeech}, 2021,
  pp. 2561--2565.

\bibitem{xu2021convolutional}
H.~Xu, Y.~Huang, Y.~Zhu, K.~Audhkhasi, and B.~Ramabhadran, ``Convolutional
  dropout and wordpiece augmentation for end-to-end speech recognition,'' in
  \emph{ICASSP}, 2021.

\bibitem{yu2021fastemit}
J.~Yu, C.-C. Chiu, B.~Li, S.-y. Chang, T.~N. Sainath, Y.~He, A.~Narayanan,
  W.~Han, A.~Gulati, Y.~Wu \emph{et~al.}, ``{FastEmit}: Low-latency streaming
  {ASR} with sequence-level emission regularization,'' in \emph{ICASSP}, 2021.

\bibitem{mahadeokar2021alignment}
J.~Mahadeokar, Y.~Shangguan, D.~Le, G.~Keren, H.~Su, T.~Le, C.-F. Yeh,
  C.~Fuegen, and M.~L. Seltzer, ``Alignment restricted streaming recurrent
  neural network transducer,'' in \emph{Spoken Language Technology Workshop
  (SLT)}, 2021.

\bibitem{hu2020exploring}
H.~Hu, R.~Zhao, J.~Li, L.~Lu, and Y.~Gong, ``Exploring pre-training with
  alignments for {RNN} transducer based end-to-end speech recognition,'' in
  \emph{ICASSP}, 2020.

\bibitem{zhang2020transformer}
Q.~Zhang, H.~Lu, H.~Sak, A.~Tripathi, E.~McDermott, S.~Koo, and S.~Kumar,
  ``Transformer transducer: A streamable speech recognition model with
  transformer encoders and {RNN-T} loss,'' in \emph{ICASSP}, 2020.

\bibitem{han2020contextnet}
W.~Han, Z.~Zhang, Y.~Zhang, J.~Yu, C.-C. Chiu, J.~Qin, A.~Gulati, R.~Pang, and
  Y.~Wu, ``Contextnet: Improving convolutional neural networks for automatic
  speech recognition with global context,'' \emph{arXiv:2005.03191}, 2020.

\bibitem{gulati2020conformer}
A.~Gulati, J.~Qin, C.-C. Chiu, N.~Parmar, Y.~Zhang, J.~Yu, W.~Han, S.~Wang,
  Z.~Zhang, Y.~Wu \emph{et~al.}, ``Conformer: Convolution-augmented transformer
  for speech recognition,'' \emph{arXiv:2005.08100}, 2020.

\bibitem{radfarconvrnn}
M.~Radfar, R.~Barnwal, R.~V. Swaminathan, F.-J. Chang, G.~P. Strimel,
  N.~Susanj, and A.~Mouchtaris, ``{ConvRNN-T}: Convolutional augmented
  recurrent neural network transducers for streaming speech recognition.''

\bibitem{kim2020accelerating}
J.~Kim and Y.~Lee, ``Accelerating rnn transducer inference via one-step
  constrained beam search,'' \emph{arXiv preprint arXiv:2002.03577}, 2020.

\bibitem{kang2022fast}
W.~Kang, L.~Guo, F.~Kuang, L.~Lin, M.~Luo, Z.~Yao, X.~Yang, P.~{\.Z}elasko, and
  D.~Povey, ``Fast and parallel decoding for transducer,'' \emph{arXiv preprint
  arXiv:2211.00484}, 2022.

\bibitem{sennrich2015neural}
R.~Sennrich, B.~Haddow, and A.~Birch, ``Neural machine translation of rare
  words with subword units,'' \emph{arXiv:1508.07909}, 2015.

\bibitem{panayotov2015librispeech}
V.~Panayotov, G.~Chen, D.~Povey, and S.~Khudanpur, ``Librispeech: an asr corpus
  based on public domain audio books,'' in \emph{2015 IEEE international
  conference on acoustics, speech and signal processing (ICASSP)}.\hskip 1em
  plus 0.5em minus 0.4em\relax IEEE, 2015, pp. 5206--5210.

\bibitem{wang2021voxpopuli}
C.~Wang, M.~Riviere, A.~Lee, A.~Wu, C.~Talnikar, D.~Haziza, M.~Williamson,
  J.~Pino, and E.~Dupoux, ``{VoxPopuli}: A large-scale multilingual speech
  corpus for representation learning, semi-supervised learning and
  interpretation,'' \emph{arXiv:2101.00390}, 2021.

\bibitem{pratap2020mls}
V.~Pratap, Q.~Xu, A.~Sriram, G.~Synnaeve, and R.~Collobert, ``{MLS}: A
  large-scale multilingual dataset for speech research,''
  \emph{arXiv:2012.03411}, 2020.

\end{thebibliography}
\end{document}